# Neutral particle release from Europa's surface

C. Plainaki[(1)], A. Milillo[(1)], A. Mura[(1)], S. Orsini[(1)] and T. Cassidy[(2)]

[(1)] INAF - Istituto di Fisica dello Spazio Interplanetario Via del Fosso del Cavaliere, 00133 Roma, Italy

[(2)] University of Virginia, USA

Number of pages: 42

Number of Figures: 9

Number of Tables: 1

proposed running head:

*Plainaki et al., 2009: Neutral particle release from Europa*

**Abstract**

In this paper, we look at space weathering processes on the icy surface of Jupiter's moon Europa. The heavy energetic ions of the Jovian plasma ($H^+$, $O^+$, $S^+$, $C^+$) can erode the surface of Europa via ion sputtering (IS), ejecting up to 1000 $H_2O$ molecules per ion. UV Photons impinging the Europa's surface can also result in neutral atom release via photon-stimulated desorption (PSD) and chemical change (photolysis). In this work, we study the efficiency of the IS and PSD processes for ejecting water molecules, simulating the resulting neutral $H_2O$ density. We also estimate the contribution to the total neutral atom release by the Ion Backscattering (IBS) process. Moreover, we estimate the possibility of detecting the sputtered high energy atoms, in order to distinguish the action of the IS process from other surface release mechanisms. Our main results are: 1) The most significant sputtered-particle flux and the largest contribution to the neutral $H_2O$-density come from the incident $S^+$ ions; 2) The $H_2O$ density produced via PSD is lower than that due to sputtering by ~1.5 orders of magnitude; 3) In the energy range below 1 keV, the IBS can be considered negligible for the production of neutrals, whereas in the higher energy range it becomes the dominant neutral emission mechanism; 4) the total sputtering rate for Europa is $2.0 \cdot 10^{27}$ $H_2O$ $s^{-1}$; 5) the fraction of escaping $H_2O$ via IS is 22% of the total sputtered population, while the escape fraction for $H_2O$ produced by PSD is 30% of the total PSD population. Since the PSD exosphere is lower than the IS one, the major agent for Europa's surface erosion is IS on both the non-illuminated and illuminated side. Lastly, the exospheric neutral density, estimated from the Galileo electron density measurements appears to be higher than that calculated for $H_2O$ alone; this favours the scenario of the presence of $O_2$ produced by radiolysis and photolysis.

Keywords: EUROPA, SATELLITES, SURFACES.

# 1. Introduction

The Jovian System is a small planetary system, built-up out of the mixture of gas and material that was present in the outer regions of the solar nebula. Through a complex history of accretion, internal differentiation and dynamic interaction, a unique satellite system was formed. Among them, Europa is believed to shelter an ocean between its geo-dynamically active icy crust and its silicate mantle. Europa occupies a transitional position between Io and the icy satellites Ganymede and Callisto. Its surface is very young; both crater counting and estimation of erosion rates by sputtering (constrained by the measurements of $H_2O$ escape rates by Galileo) suggest an age of Europa's surface between 10 million and one billion years.

The radiation environment of Europa includes fluxes of energetic electrons, protons, $O^{n+}$, $S^{n+}$, $H^+$ solar photons, and meteoroids (Cooper et al., 2001). As a consequence of the existence of such an environment, the surface of Europa can be modified both physically (via sputtering, i.e., the erosion of a surface) and chemically (via radiolysis, the chemistry that follows the impact of ionising radiation). Sputtering, typically by ions (IS), and Photon-stimulated Desorption (PSD), by photons, are commonly thought of as two of the leading processes that drive neutrals from an icy surface. Because of the fact that dissociation and chemical reactions produced by ionizing radiation also release energy that can cause the ejection of atoms or molecules, sputtering and radiolysis are closely related. Although there is a number of review articles on these processes (Strazzulla, 1998; Johnson, 1998; Madey et al., 2002; Baragiola, 2003; Johnson et al., 2003; 2009) the behaviour of the materials and the exact processes taking place at the temperatures at Europa have not been understood. However, the surface erosion rate depends on the efficacy of all processes that take place on the satellite's surface, as well as on the availability of fresh ice to replace that lost to space. Measurements of the incoming/reflected radiation and particle fluxes can help identify the structure and the composition of Europa's exosphere and surface. They are also useful, for establishing the net production, loss, and exchange of key chemical species between the magnetosphere, exosphere, surface and subsurface.

In this work, we attempt to quantify the neutral particle fluxes emerging from the icy surface of Europa. The paper is organized as follows. Section 2 presents a new Monte Carlo (MC) model, for demonstrating the neutral-particle imaging of the plasma-surface interaction. Specifically, in section 2.1 an overview of the Europa's surface composition is presented. In section 2.2 the main neutral release processes from a planetary body are discussed. Thermal Desorption (TD), Electron-stimulated Desorption (ESD) and Micrometeoroid Impact Vaporization (MIV) are discussed while IS, Ion Back-scattering (IBS) and PSD are analysed each one separately in Sections 2.2.1, 2.2.2 and 2.2.3 respectively. In section 2.3 the numerical procedure is presented. In Section 3, the results of the simulations are presented and discussed. Finally, in Section 4, the main conclusions of this work are summarised.

## 2. A model for the neutral particle generation by the surface of Europa
### 2.1. Europa's surface composition

The average density of Europa, equal to ~2989 kg/m$^3$ (Anderson et al., 1998), is close to that of the Moon (~3340 kg/m$^3$), suggesting a rock-dominated Europa composition. However, the earliest telescopic and spectroscopic observations suggested that Europa's surface is primarily covered by $H_2O$ ice with an admixture of trace species and it has the highest albedo of the Galilean satellites (Kuskov and Kronrod, 2005). The surface of Europa is relatively smooth and geologically young; crater preservation yields an age estimate of ∼10–100 Ma (Moore et al., 1998). Although the $H_2O$ is the dominant species on the surface of Europa (>99% according to Clark, 2004), there are also traces of non-icy material: $H_2O_2$, with a surface concentration of about 0.13 % by number of molecules, trapped $SO_2$ and $CO_2$ (Tiscareno and Geissler, 2003; McCord et al, 1998) and some mixture of hydrated $H_2SO_4$ and sulfate salts (Orlando et al., 2005). A still open question is whether these minor species are native to the satellite or are synthesized by exogenic processes such as ion irradiation (Carlson, 2002; 2004). In particular it has been suggested that $SO_2$ could have been formed by magnetospheric sulphur implanted in $H_2O$ ice (Lane et al. 1981; Eviatar et al., 1985; Sack et al. 1992). Similarly, it has been suggested that part of the observed $CO_2$ could be due to implantation of C ions in $H_2O$ ice (Strazzulla et al. 2003). Redeposition of sputter products (Johnson et al. 1988; Cassidy et al. 2009) can also affect the spatial distribution of the trace species.

**2.2.    Neutral emissions from the surface**

Important neutral release processes from planetary surfaces that are generally active on the Solar System (SS) bodies are a) IS, b) IBS, c) PSD, d) TD, e) ESD and f) MIV.

The IS process, active, for instance, at Mercury (e.g., Milillo et al., 2005) and the Moon (e.g., Wurz et al., 2007), is defined as the removal of atoms or molecules from a solid surface, due to the interaction of a projectile ion with target electrons and nuclei (Johnson, 1990). The struck nuclei in turn produce a cascade of collisions between target atoms leading to additional ejecta (Sigmund, 1981). IS is also an important release process for the case of icy objects embedded in planetary magnetospheres (e.g. Lanzerotti et al., 1978). At Europa, it constitutes an efficient process of erosion, due to the high rate of bombardment by ions from the Io plasma torus (Lane et al., 1981), together with the satellite's easily sputtered icy surface.

Sputtering and subsequent erosion occur mostly when the moon's atmosphere cannot impede the precipitating ion from reaching the surface or the liberated neutrals from escaping (Hall et al., 1995). Various models for Europa's atmosphere, based on observations and theoretical considerations, suggest that Europa may have an atmosphere that originates from frozen surface water ice. The first observations of Europa atmosphere were made by Hall et al. (1995) using the Hubble Space Telescope. Based on the ratio of the atomic oxygen line intensities at 1304Å and 1356Å, they proposed that these emissions resulted from dissociative excitation of $O_2$. Moreover, the observations from Cassini suggested $O_2$ emission near the surface and a tenuous extended atomic O component (Hansen et al., 2005). The likely main atmospheric constituents of Europa's atmosphere are $H_2O$, $O_2$, O, $H_2$, OH and H. Assuming $O_2^+$ is the dominant ion and dissociative recombination is the dominant loss process, Kliore et al. (1997) calculated the $O_2$ density (at the surface) required to produce the electron density observed by the Galileo spacecraft, as $\sim 3 \cdot 10^{14}$ m$^{-3}$. If the electron impact ionization were considered, these authors found that an even lower molecular oxygen density would be consistent with their data ($\sim 5 \cdot 10^{13}$ m$^{-3}$). Replacing $O_2$ by any of the other likely constituent would not significantly change the resulting density at surface, since neither the dissociative recombination rate coefficients nor the ionization efficiencies are significantly different. For example, if $H_2O$ is taken to be the dominant neutral molecular species, which leads to $H_3O^+$ as the main ion species, the required neutral density is still on the order of $10^{14}$ m$^{-3}$ (Kliore et al., 1997). Shematovich et al. (2005) modelled Europa's atmosphere using sputtering and radiolysis data and found, at an orbital altitude of 100 km, densities of $10^{12}$ to $10^{13}$ m$^{-3}$ for $O_2$, and $10^{10}$ to $10^{12}$ m$^{-3}$ for $H_2O$ and O. Moving to a lower orbit 10 km would increase the $O_2$ and $H_2O$ densities by an order of magnitude but would negligibly change the O and OH densities. Cassidy et al. (2009) using the results of earlier simulations, have modelled the atmospheric spatial structure of the volatile species $CO_2$ and $SO_2$. They found that a neutral mass spectrometer orbiting at about 100 km above the surface could detect Europa's atmosphere species with surface concentrations above about 0.03%, providing important information on its atmosphere. On the basis of the neutral atmospheric densities estimated by Kliore et al. (1997), in this work, we estimate a range of the mean free paths for $O_2$ in Europa's atmosphere from 13 km to 78 km. We also estimate the scale height, on the basis of formula $H=k_BT/(Mg)$, where $k_B$ is the Boltzman's constant, $T$ is Europa's temperature at the surface, $M$ is the molecular weight of $O_2$ and $g$ is the acceleration due to gravity and we find that it varies from 17 km to ~26 km, in good agreement with the estimation of 20 km given by Ip (1996). As a first approximation, therefore, we assume a collision-less atmosphere, where the surface ion sputtering is possible.

Johnson (1990) did in fact set an upper limit of the column density so that the ions are able to reach the surface where they produce sputtering. Other authors have recently given larger values for the scale height, such as 135 km (Leblanc et al., 2002) calculated, however, on the basis of the Na portion of the atmosphere which has far lower densities than either $O_2$ or $H_2O$. Due to the fact that the issue of a collision-less atmosphere of Europa is rendered quite complicated, we can conclude that, within the perhaps poor constraints, the mean free path and the scale height are of the same order of magnitude. Consequently, as a first approximation we can consider that the atmosphere is not an obstacle to the travel of the impinging ions as well as to the escape of the sputtered particles. In this condition processes generating neutral atoms from the atmosphere, such as charge exchange and atmospheric sputtering, can be considered negligible. A second approach, including an analytical study of the role of collision regimes at Europa's atmosphere is intended in the near future.

When ions impinge on a surface, Ion Back Scattering (IBS) can take place; nevertheless IBS is not a surface release process. Since impacting ions with energy less than about 1 MeV will interact with the surface as neutrals, the IBS process, at a first approximation, can be treated as a mechanical elastic collision between impinging ions and target surface molecules.

PSD refers to the desorption of neutrals or ions as a result of direct excitation of surface atoms by an incident photon (Hurych, 1988). For Europa, the interest in the effects of ultraviolet radiation on $H_2O$ ice arises both from its significance as an analogue for the damage of biological materials as well as from the need to understand the stability of ice grains or ice mantles on refractory grains in the interstellar medium (Sternberg et al., 1987) and in the SS (Carlson, 1980).

In Thermal Desorption (TD) molecules are released into gas phase due to the heat content of the surface. There are two types of binding to the surface: chemisorption and physisorption. Chemisorption involves the actual formation of bonds between the adsorbate and the substrate whereas the second type of adsorption, involves van der Waals interactions such as dispersion forces or a dipolar interaction. On the basis of the Galileo observations, Europa's temperature ranges between 86 K and 132 K (Spencer et al., 1999), corresponding to a thermal energy range of from 0.011 eV to 0.017eV. Since the surface molecules have a thermal distribution, the vapour pressure produces the sublimation. For disk-averaged temperature of 106K, on the basis of the equations given in Spencer (1987), a vapour pressure of $10^{-16}$ Pa and a sublimation rate of $10^{11}$ $H_2O$ $m^{-2}$ $s^{-1}$ is estimated (Shematovich et al., 2005). However, locally, at sites where the temperature is relatively high, the $H_2O$ vapour pressure and the sublimation rate increase (for 132K they become of the order of $10^{-12}$ Pa and $10^{15}$ $H_2O$ $m^{-2}$ $s^{-1}$ respectively). Therefore this process cannot be considered negligible everywhere.

The ESD refers to the surface neutral release process via electron impact. The flux released via this process is proportional to the impacting electron flux, the ESD cross-section, and Europa's surface numerical density ($10^{19}$ $m^{-2}$). Considering that the energy and angle integrated electron flux impacting Europa is $10^{12}$ $m^{-2}s^{-1}$ (after Paranicas et al. 2001) and that the ESD cross section is of the order $10^{-15}$-$10^{-13}$ $m^2$, the estimated $H_2O$ release flux is not higher than $10^{10}$ $m^{-2}$ $s^{-1}$.

The MIV refers to the impact vaporization caused by micrometeorites hitting the surface of a planet. According to Cooper et al. (2001), the energy input from meteoroids is of several orders of magnitude less than the particle energy fluxes and as a result the meteoroid energy effects, although they may act locally, are not significant for vapor production on any of the Galilean satellites. On the basis of the formulas contained in Cintala (1992), we make a rough estimation of the expected $H_2O$ vapour flux produced by small meteoroids impacting on the surface of Europa. Considering an interplanetary meteoroid mass flux **of $7 \cdot 10^{-12}$ g $s^{-1}$ $m^{-2}$ in the region of Europa (Krüger et al., 2003),** and a **maximum velocity for the impacting meteoroids** equal to their jovian escape velocity near the orbit of Europa (i.e. 19km/s after Cooper et al., 2001) and assuming that the projectile material is diabase (Cremonese et al., 2005), we estimate a vapour $H_20$ flux equal to **$3.8 \cdot 10^{11}$ $m^{-2}$ $s^{-1}$.**

The particles released from Europa surface are either lost in space or they return to the surface if their energy is less than that corresponding to the escape velocity ($E_{esc}$= 0.38 eV). All neutral particles escaping from the moon's gravity are assumed either to remain on a ballistic trajectory or to be ionized.

**2.2.1. Ion-sputtering study**

IS is the ejection of molecules from a surface due to bombardment by energetic charged particles. It can remove up to thousands of $H_2O$ molecules per incident ion **(Shi et al., 1995)**, which then follow ballistic trajectories until they either redeposit onto the surface, become part of the bound atmosphere, or escape from the gravity field or are ionized and picked up. The ambient gas produced by sputtering (Johnson, 1990; 1998) depends on: a) the composition and the chemical structure of the surface, reproducing more or less its local composition on atomic level (Wurz et al., 2007; Johnson, 1990; 1998), b) the ion-radiation incident on the satellite, and c) the yields, i.e. the number of liberated particles per incident ion. These sputtering yields are roughly uniform over Europa's surface (Paranicas et al., 2002). Preferential sputtering of the different elements of a compound can lead to a surface enrichment of those elements with low sputtering yields. However this seems not to be the case in Europa, since the sputtering yields are large resulting in trace species being carried off with the $H_2O$ molecules (Cassidy et al., 2009). Tiscareno and Geissler (2003) argue that the number of re-depositing particles cannot overcome sputtering erosion, due to gravitational escape and removal by electron impact ionization. Therefore, the global average result of sputtering is net erosion.

In this model we study the case that only $H_2O$ molecules are being sputtered from Europa, by the Jupiter magnetospheric plasma ions. Possible ejection of trace species such as sodium (Cipriani et al., 2008) have not been taken into account. The physical motivation for this assumption is that trace species concentrations are thought to be small: e.g., sodium surface concentration is ~ 0.8% (Johnson, 2000; Leblanc et al., 2005, Cassidy et al., 2009).

*Impinging particle fluxes*

In order to simulate the flux of the impinging particles we integrated over the $H^+$, $O^+$ and $S^+$ ion differential fluxes presented in Paranicas et al. (2002). For the calculation of the $C^+$ ion flux we assumed the value given by Strazzulla et al. (2003), estimated by using C/S ratios given by Hamilton et al. (1981) (see Table-I). As stated earlier, the ejected particles are all assumed to be $H_2O$-molecules. We produced results for the flux and the density distribution of the emerging particles separately for each type of ion irradiation contained in Table-I.

*Sputtering Yields*

The sputtering yield is defined as the number of molecules ejected by a single ion impinging on a surface. Here we use ion-induced sputtering yields for "H-like" and "O-like" ions compiled by Shi et al. (1995) from laboratory measurements for different temperatures and different impinging ion energies. Above ~60K water ice sputtering yields increase with temperature (Johnson et al. 1990; Shi et al., 1995; Famà et al. 2008). Based on such results, and taking into consideration the fact that the Galileo observations of Europa's thermal emission showed that low-latitude diurnal brightness temperatures range between 86 to 132 K (Spencer et al., 1999), we calculated the sputtering yield for $H_2O$ for incident $H^+$ and $O^+$ using the following equation:

$$Y(T, E_i) = Y_0(E_i) + Y_1(E_i) e^{-E_a(E_i)/k_B T} \qquad (1)$$

Here $k_B$ is the Boltzman's constant, $T$ is the temperature of Europa, assumed to be ~106 K (Spencer et al., 1999), and parameters $Y_0$, $Y_1$, $E_\alpha$ take values that depend on the type and the energy, $E_i$, of the impinging ion, i.e.: *$Y_0$=2.76, $Y_1$ =880, $E_a$=0.052eV* for 30keV $H^+$ and *$Y_0$=37.4, $Y_1$ =40000, $E_a$=0.074eV* for 30keV $O^+$ (Shi et al., 1995). Consequently, for 30 keV ions, the yields take the values 6 for $H^+$ and 50 for $O^+$. Such a difference between the two sputtering yields is quite reasonable because the oxygen ions have a relatively higher energy deposition per unit depth inside the surface, i.e. higher stopping power (*dE/dx*). Similar yield values of the various species are obtained when the formula given in Johnson et al. (2004) or Fama et al. (2008) are used. The temperature dependence of the sputtering yield is caused by reactions between radicals for which trapping, diffusion and reaction rates are temperature dependent (Boring et al., 1983; Reinmann et al., 1984; Shi et al., 1995). In this model, we simplify the calculation procedure assuming a uniform temperature of the surface of Europa. However, the temperature is not uniform; a temperature map, based on the Galileo mission data, has been given by Spencer et al. (1999). These data were further fit to an analytical function in order to provide a precise latitude and solar zenith angle dependent temperature map (Cassidy et al., 2007). This non-uniform temperature, which influences the sputtering yield (equation (1)), will be considered in a future study.

Here we also assume that the sputtering rate is spatially uniform, as most of the sputtering on Europa is done the energetic heavy ions (sulfur and oxygen) which impact the surface nearly uniformly (Cassidy et al., 2009 Cooper et al., 2001; Paranicas et al., 2002).

*Binding energies*

The energy with which a molecule is ejected from the surface is affected by the surface binding energy and not by the energy or identity of the impacting ion (Johnson, 1990; 1998). Although the sublimation energy of $H_2O$ is 0.45 eV/molecule, the sputtered particle energy distributions for molecular ices tend to have maxima at lower energies than a collision cascade prediction with surface binding energy equal to the normal sublimation energy (Brown and Johnson, 1986; Boring et al., 1983; Brown et al., 1984; Haring et al., 1984). Several explanations for this phenomenon have been proposed; the surface may be strongly disrupted with many atoms or molecules leaving at once without experiencing the same binding energy as a single atom leaving a planar surface (Roosendaal et al., 1982; Reimann et al., 1984). In addition, the surface region may be electronically and collisionally excited and the interatomic or intermolecular forces are lower as a result of that excitation (Reimann et al., 1984; Watson et al., 1983). In this study we perform a sputtering simulation that corresponds to an 'effective' binding energy $E_b$=0.054 eV, which was experimentally obtained in the past (Boring et al. 1984, Haring et al., 1984).

*Energies and directions of sputtered particles*

The energy distribution of the ejected molecules affords vital information about the fate of the sputtered molecules. In this paper we used a form for the energy-distribution given by Sigmund (1969) for knock-on sputtering:

$$f_s(E_e, E_i) = \begin{cases} \sim \dfrac{E_e}{(E_e + E_b)^3} \left[ 1 - \left( \dfrac{E_e + E_b}{T_m} \right)^{1/2} \right] & E_e \leq E_i - E_b \\ 0 & E_e > E_i - E_b \end{cases} \qquad (2)$$

where $E_e$ is the ejection energy, $E_b$, is the 'effective' binding energy and $T_m$ is the maximum energy transferred in the collision and **$E_i$ is the energy of the impacting particle before the collision.** The shape of the IS energy-distribution function, for 100 keV S$^+$ (32 amu) and ejected H$_2$O molecules is shown in Fig.1. The 'effective' surface binding energy $E_b$ influences significantly this energy spectrum. In Fig. 1 we have considered $E_b$ = 0.054 eV (Boring et al., 1984). The fraction of sputtered escaping particles ($E > E_{esc}$) is about 22%.

The polar angle distribution of sputtered atoms, is generally described by a $cos^k(\varphi_e)$ law (Hofer, 1991), where the exponent $k$ depends on the structure of the surface and $\varphi_e$ is the ejection angle relative to the normal. In our model, for the exponent $k$, we have used the assumption of Cassidy and Johnson (2005) suitable for the fine-grained and porous regolith, hence *k=1*. For the azimuth angle we used a uniform distribution over $2\pi$. A summary of the sputtering-input parameters, used in our program (sputtering yield, relative abundance, 'effective' binding energy, etch.) is presented in Table-I. The columns of the incident flux values refer only to the energetic hot and non-thermal plasma.

**2.2.2. Ion back scattering study**

The high energy ions impinging upon Europa penetrate the moon's surface, to distances from some hundreds of nm up to some µm, until they entirely lose their kinetic energy. During their penetration, ions can be neutralized and make close collision with target molecules of the surface. As a result, a small fraction of the initial ion flux can be scattered back out as of the surface, as neutral. In this paper, we refer to this process as IBS.

When a particle is scattered elastically from a target atom at an angle $\theta$, the ratio of the scattered-ion energy $E$ to the incident energy, $E_i$, can be calculated from the conservation of energy and momentum. The so called kinematic factor in this case is given by the following equation:

$$K = \frac{E}{E_i} = \left[ \frac{\sqrt{M_2^2 - M_1^2 \sin^2\theta} + M_1 \cos\theta}{M_1 + M_2} \right]^2 \qquad (3)$$

where $M_1$, $M_2$ are the masses of the incident ion and target atom, respectively and $\theta$ is the scattering angle (i.e. the angle between the velocity of the impinging particle and that of the scattered one). Back-scattering exists for cases where $\theta \geq 90^0$. The resulting energy of the scattered ion decreases with increasing scattering angle (see Fig. 2). From Fig. 2 it is derived that the IBS is more efficient for the lighter ions. Therefore, not all of the incident ion species are back-scattered by the $H_2O$ molecules of Europa's surface with the same efficiency. The interactions between H and $H_2O$ molecules, as well as those between C and $H_2O$ molecules, are the most efficient in causing the back-scattered of the impacting species. Due to the fact that the kinematic factor of C – $H_2O$ collisions, is much smaller than that of H – $H_2O$ collisions (see Fig. 2), in the following we focus on the H – $H_2O$ collisions.

The efficiency of this process depends also from the surface type and from the ion's impact energy, but the exact value is not easily derived. Recently, ENA measurements from the Moon resulting from solar wind backscattering (Mc Comas et al. 2009; Wieser et al. 2009) demonstrated that the neutrals coming out from IBS are about 10-20% of the impinging ion-flux; moreover, the ionized part of the backscattered material is only few percent of the IBS neutral fraction (Saito et al, 2008). However the above measurements refer to 1-keV protons impinging onto a regolith surface. In Europa's case the impinging protons are far more energetic (10-100 keV) and the surface is iced, hence the application of the same yield could be not accurate. Fig. 3 shows the energy spectrum of the 10-keV protons back-scattered from an icy surface, as calculated with TRIM/SRIM software (Ziegler et al., 1988). It is clearly seen that the normalized yield is of the order of $10^{-4}$ for ejecting energies up to 1 keV, whereas for more energetic particles it becomes smaller. **However, the IBS mechanism is still an open question and more theoretical and experimental studies are necessary in this field. Finally, future missions will offer the possibility to have in situ IBS measurements and will reveal the role of this process in the surface ageing.**

### 2.2.3. Photon-stimulated desorption study

PSD primarily ejects volatiles such as Na, K (Madey et al., 1998;2002) but it has also been shown to eject $H_2O$ (Westley et al. 1995). The physical mechanism varies for different adsorbate/substrate systems, and is either a direct or an indirect photon-induced electronic excitation of a surface atom (Zhou et al., 1991; Madey et al., 1998). It has been assumed that desorption by background UV photons is an important mechanism for low temperature ices in space. Westley et al. (1995) have measured the absolute yield for photon desorption of low temperature $H_2O$ ice (35-100 K) by Lyman-$\alpha$ (10.2 eV) photons.

The surface of Europa that is in sunlight is exposed to a flux of photons, some of which can lead to desorption neutrals. The neutral flux via PSD of the icy surface of Europa can be calculated on the basis of the following equation (Wurz and Lammer, 2003):

$$f_{PSD} \approx \frac{1}{4} f_{photon} \sigma_{PSD} F d \cos(\phi) \qquad (4)$$

where $f_{photon}$ is the photon flux per unit area per time, integrated over the photon energy range ≥ 7eV and equal to $5.8 \cdot 10^{14}$ s$^{-1}$m$^{-2}$ in the vicinity of Europa, $F$ is the fraction of H$_2$O ice on the surface, equal to 0.99, $d$ is the numerical surface density and $\varphi$ is the location of one point on the surface. The measured yield from low temperature (140 K) poly-crystalline water ice under irradiation with Lyman-α (121.6 nm) is $Y_{psd} = 3.8 \; 10^{-3}$ (Westley et al. 1995). On the basis of these considerations we find a PSD cross section equal to $3.5 \; 10^{-22}$ m$^2$, which is higher to that corresponding to rough regolith surfaces (i.e. ranging from $1.0 \cdot 10^{-24}$ m$^2$ to $1.0 \cdot 10^{-23}$ m$^2$ according to Yakshinskiy and Madey (1999)). In this model, the dependence of the PSD of ice on the radiation dose (Westley et al., 1995) has not been considered. In a recent theoretical work by Grigorieva et al. (2007), the PSD yields for ice were estimated for different temperatures, being $Y = 3 \times 10^{-3}$ for T ≤ 35 K and $Y = 7 \times 10^{-3}$ for T ≥ 100 K. It is worth noticing that both these independently derived ranges for Y are comfortably of the same order of magnitude.

For the ejected atoms, a Maxwellian distribution has been considered (Yabushita et al., 2009):

$$f_{PSD} = \frac{E_e}{(k_B T_e)^2} \exp\left(-\frac{E_e}{k_B T_e}\right) \tag{5}$$

where $E_e$ is the emerging particle energy, $k_B$ is the Boltzman's constant and $T_e$ is an effective temperature for the process, equal to 1800K (Yabushita et al., 2009). The form of the energy distribution function, for this temperature is presented in Fig. 4. It is clearly seen that this distribution shows a peak at $E_e \sim 0.17$ eV. The fraction of escaping particles ($E_e >E_{esc}$) is about 30% of those escaping via the PSD mechanism.

## 2.3. Numerical model procedure

In order to study the IS and the PSD effect taking place on the icy surface of Europa and, more specifically, to calculate the space flux and the space density of the ejected sputtered H$_2$O particles, we applied a Monte Carlo (MC) method over a cubic space of $R_m$ x $R_m$ x $R_m$ where $R_m$ is set equal to the five times Europa's radius ($R_m = 5R_E$). In this space Europa is positioned at the origin. The space inside which the MC method is carried out is enclosed by the Hill sphere (Hill's radius ~ 8.7 $R_E$). The number of test particles is set equal to 100000. In the beginning each particle is placed at a random location on Europa's surface given a random velocity and angular distribution according to those described in previous section. Then the particle trajectories are integrated taking into account the gravitational fields of both Europa and Jupiter, as well as Europa's rotation. The integration time step is set equal to $dt = \dfrac{dx}{\upsilon_{nor} \cdot n_c}$, where $n_c$ is the number of particles inside a cell, $\upsilon_{nor} = \sqrt{\dfrac{2E_e}{m_e}}$ is the velocity of the emerging particle and $dx = \dfrac{2R_m}{n}$ is the dimension of each cell ($n$ is the number of bins for each dimension, set to 100).

*Gravitational Forces*

With Europa at the origin and Jupiter at $x = -D_E$ (where $D_E$ is Jupiter's distance from Europa equal to $6.7 \cdot 10^8$ m), the gravity force upon a particle emerging from Europa's surface is given by:

$$\vec{F_G} = -\dfrac{GM_E}{r^3}\vec{r} - \dfrac{GM_J}{r'^3}\vec{r'} \tag{6}$$

where $M_E, M_J$ are the masses of Europa and Jupiter respectively, $\vec{r} = \left[x\hat{i}, yj, zk\right]$ is the particle's position vector in respect to Europa and $\vec{r'} = \left[(x+D_E)\hat{i}, yj, zk\right]$ is the particle's position vector in respect to Jupiter.

*Europa's rotation*

Given the angular velocity of the rotation of Europa directed along the $z$ axis (period $T_E = 3.55d$), the vector of the Coriolis acceleration of the neutral particles emerging from the planet's surface, is given by :

$$F_{Cor} = 2m\Omega[-\upsilon_y \hat{i}, \upsilon_x j] \tag{7}$$

where $\Omega$ is the angular velocity of Europa's orbit ($\Omega = \sqrt{\dfrac{GM_E}{R_E^3}}$, $R_E$ is Europa's radius), $m$ is the mass of a single particle and $\vec{\upsilon} = \left[\upsilon_x \hat{i}, \upsilon_y j, \upsilon_z k\right]$ is its velocity in the Europa rotating reference frame.

*Electron-impact ionization*

In the numerical model, the trajectories of $H_2O$ molecules sputtered from Europa's surface are followed until they either redeposit back onto the surface or are ionized and carried away by Jupiter's magnetosphere. According to Tiscareno and Geissler (2003) the number of $H_2O$ molecules lost by in-flight ionization can cause the number of re-impacting $H_2O$ molecules to decrease by 10%. Electron-impact ionization is accounted stochastically by assigning at each simulated particle a lifetime, with a probability following the Poisson distribution, before it is launched from Europa's surface. If this lifetime expires before the molecule re-impacts the surface, the molecule is considered to be ionized and is removed from the model. We assume, as a first approximation, that the electron environment near Europa is fairly homogeneous, although a hemispherical difference (major in the trailing hemisphere) in energetic electrons' energy deposition does exist (Paranicas et al., 2001). This may affect the present results by at least a factor of two. The ionization lifetime for $H_2O$ molecules in Europa's vicinity is calculated equal to 51.2 hours (Tiscareno and Geissler, 2003).

In this study, the photo-ionization is considered negligible.

## 3. Model Results -Discussion

The sputtered particle total fluxes and densities produced separately by the $H^+$, $S^+$, $O^+$ and $C^+$ impinging particles are presented in Fig. 5, for $E_b$=0.054 eV. . The most significant emerging flux comes from $S^+$ impinging the Europa's surface. The sputtered particle flux at the surface, in case of impinging $S^+$, is calculated to be ~ $2.1\cdot 10^{13}$/s/m$^2$, that is bigger than the other considered cases. The respective sputtered particle density at the surface is calculated to be ~ $1.6\cdot 10^{10}$/m$^3$. The total flux of sputtered $H_2O$ molecules, produced at the surface by all types of impinging ions, is calculated to be $3.2\cdot 10^{13}$ $H_2O$/s/m$^2$ and the total density $2.7\cdot 10^{10}$ $H_2O$/m$^3$. These results are in general in good agreement with calculations realized by other studies. For example, Ip et al. (1998), on the basis of the data from the Energetic Particle Detector (EPD) experiment on the Galileo mission, have shown that the sputtering rate per unit area of protons, sulphur and oxygen ions are $Q(H^+)=4.1\cdot 10^{11}$ $H_2O$/s/m$^2$, $Q(S^+)=8.5\cdot 10^{12}$/s/m$^2$ and $Q(O^+)=7.9\cdot 10^{12}$/s/m$^2$ respectively. These fluxes are quite close to ours: $Q(H^+)=2.0\cdot 10^{12}$ $H_2O$/s/m$^2$, $Q(S^+)=2.1\cdot 10^{13}$/s/m$^2$ and $Q(O^+)=8.9\cdot 10^{12}$ $H_2O$/s/m$^2$. The small differences in the released fluxes are due to the different corresponding impinging fluxes considered. It should be noted that the impinging ion flux, depending on external particle circulation, varies with time and space, being able to produce also higher fluxes of sputtered particles. Specifically, since at Jupiter the plasma co-rotates with the planet, it overtakes Europa continuously in its orbital motion resulting in a preferential plasma flow on the trailing hemisphere. Consequently, the higher energy particles bombard the satellite in more complex ways since their instantaneous velocities can differ from those consisting the bulk (Johnson et al., 2009). So the results of this model should be considered as a first approach to the sputtered particle flux expected around Europa, whereas a more sophisticated version of the model, taking into consideration, possible preferential plasma flow, is intended in the future.

    On the basis of the sputtered particle fluxes calculated by our model, we estimate the total sputtering rate for Europa, taking into account the finite gyroradius effect of the energetic ions (Pospieszalska and Johnson, 1989). We found that the sputtered particle rate is $2.0\cdot 10^{27}$ $H_2O$ s$^{-1}$, a result that is, in fact, close to those derived by other researchers (Cooper et al., 2001; Paranicas et al., 2002). It should be underlined, however, that our calculation is based on a probably overestimated sputtered particle flux emerging globally from the planet surface and hence the total sputtering rate may exhibit variations. Moreover, the "global" sputtering rates of

Cooper et al. (2001) and Paranicas et al. (2002) are inferred from upstream omnidirectional measurements and not from modelling of incident fluxes across the whole surface. Since the average surface flux would likely be respectively higher and lower than the upstream flux on the trailing and leading hemispheres, the upstream flux (π times the directional flux) gives an approximate average. In addition, the fact that the above mentioned (and subsequent) works used different set of measurements for the incident ion spectra, could lead to substantial variations in integrated primary ion and sputtered neutral fluxes. In particular, our result is in good agreement with the estimation realized by Paranicas et al. (2002) who found that the resurfacing flux in Europa due to sputtering is $8 \ 10^{13}$ s/m$^2$ which corresponds to a sputtering rate over the whole surface of $2.47 \ 10^{27}$ H$_2$O /s. Lastly, it should be mentioned that the sputtering rate extracted by analyzing the measurements of the Galileo/EPD, was $6.4 \times 10^{26}$ atoms/s (Ip et al., 1998), a result that is also close to that extracted from our model. However, one feature that is not taken into account in this model, and may influence significantly the above-mentioned results is that the surfaces of the satellites may be composed of large grains (Clark et al., 1984) with significant void space (~90%) between grains (Domingue et al., 1991). According to Cassidy and Johnson (2005) the sticking of ejecta to neighbouring grains can make the sputtering yields from a planetary surface differ considerably from the relative laboratory ones. Specifically, the H$_2$O yield from the regolith is ≈ 0.71 times that at normal incidence (Cassidy and Johnson, 2005). Therefore, in the non-ice regions of Europa, the regolith can significantly modify the relative populations of atmospheric species and their spatial distributions across the surface. If this is the case, the sputtering yields should be reduced due to sticking of sputtered species to neighbouring grains (Hapke, 1986; Johnson, 1989) and therefore lower fluxes of sputtered particles and, consequently, a lower erosion rate would be expected.

Estimations of the contribution of the secondary sputtering by pickup ions from the atmosphere have been realized in the past (for example see Ip et al. (2000)). However, the global averages calculated by these authors (up to $5.6 \cdot 10^{14}$ H$_2$O m$^{-2}$ s$^{-1}$) are heavily weighted by the maximal irradiation intensities expected for magnetospheric ions incident on the trailing hemisphere, where the ice is observed to be mixed with hydrated compounds (e.g., salts, sulphuric acid) having relatively low sputtering yield compared to pure water ice. As a result, those values may be overestimates (Shematovich et al., 2005). In addition the potential corrections for surface temperature, gyro-motion effects, and diversion of magnetospheric plasma flow by ionospheric interactions need to be considered. In this work the contribution of secondary sputtering by pickup ions is neglected due to the fact that these ions have low flux and low energy, hence low sputtering yields. Nevertheless, further model simulations will be required to resolve these discrepancies among the current models in order to address the issue, as well as more precise laboratory data in order to avoid uncertainties during calculations.

In our model, we simulate a PSD exosphere using a Maxwellian energy distribution which has been already used in numerical models of sodium exosphere in Mercury (Leblanc and Johnson, 2003; Mura et al., 2007). The $H_2O$ flux and density released via PSD (Fig. 6) reach the values of $7.2 \cdot 10^{11}$ $H_2O/s/m^2$ and $6 \cdot 10^8$ $H_2O/m^3$, respectively on the surface of the illuminated side. The emerging particle-density released via PSD is lower than that due to sputtering by about 1.5 orders of magnitude. Considering that particles that are emitted in the dayside and travel to the night side have to cross a longer path inside the atmosphere, the night-side density extracted from our model is overestimated. Depending on the mean free path assumed (ranging from 13 km to 78 km), one can estimate that only 0.3% to 3% of the particles are able to reach the night-side.

As estimated in the previous section, the IBS contribution to the bulk exosphere is negligible. Nevertheless, its role becomes more important in the higher energy part of the spectrum. In Fig. 7 the intensity versus energy spectrum of sputtered, back-scattered and PSD-ed neutrals is presented. It is clearly seen that in the energy range up to 1 keV IS is the dominant mechanism. In the higher energy range the IBS seems to be the main mechanism that produces neutrals. **Experimental and in situ IBS measurements during future missions will improve** **our understanding in this field and will reveal the role of IBS in the generation of Europa's exosphere**.

The contribution to the escape rate from Europa environment due to the two major neutral release processes can be estimated on the basis of the energy distribution function. As mentioned before the fraction of escaping particles via IS is 22% of all sputtered particles, thus meaning a total rate of $4.4 \cdot 10^{26}$ s$^{-1}$, while the fraction of escaping particles via PSD is 30% of all PSD produced particles thus meaning a total rate $3.3 \cdot 10^{24}$ s$^{-1}$. **This mean that IS dominates over PSD on both the non-illuminated and illuminated side, eroding more efficiently the surface**. According to our study, the mass erosion rate due to sputtering can be calculated by the following equation:

$$\frac{dl}{dt} = \Phi_m / d_{ice} \qquad (8)$$

where $\Phi_m$ in the sputtered flux in particles/s/m$^2$ and $d_{ice}$ in the density of ice (in kg/m$^3$). Considering that the maximum total flux of escaping sputtered H$_2$O molecules is $7 \cdot 10^{12}$ particles/s/m$^2$ and the density of ice is 931kg/m$^3$ (hence, the numerical density of ice is $\rho_{ice} N_A/M$, where $M$ is the H$_2$O molecular weight) the net erosion rate of Europa surface is calculated to be ~ 0.7m/100Myear. The values of the erosion rate of Europa's surface found in literature, vary from 6cm/100Myear (Eviatar et al., 1981) to 10m/100Myear (Johnson et al., 1981). Consequently our rough estimation seems to be inside this range. Ip et al. (1998) have calculated a net erosion rate for Europa (direct sputtering escape and ionospheric loss) equal to 20m/100Myear. This result is ~2 orders of magnitude bigger than ours but this is due to the higher sputtering rate considered by these authors. Moreover, Tiscareno and Geissler (2003) have calculated a global sputtering rate of 1.47m/100Myear whereas Cooper et al. (2001) arrived at a similar estimate of ~ 1.60m/100Myear. The use of different set of measurements for the incident ion spectra, can lead to substantial variations in the calculated sputtered neutral fluxes. Moreover, the sputtering rate of Cooper et al. (2001) was estimated for gross sputtering, so their net erosion would be of about one order of magnitude lower than 1.6m/100Myr, i.e of the same order of magnitude with the one in our estimate. It is of crucial importance to note, however, that the surface erosion effect from magnetospheric interaction might not be spatially uniform. According to Ip et al. (1998) the co-rotating thermal sulfur and oxygen atoms as well as the exospheric pick-up $O_2^+$ ion impact can render surface erosion more severe in certain localized areas.

The erosion rate due to neutral particles released via PSD and escaping the moon's gravity is estimated at 0.02m/100Myr. This result is 1.5 orders of magnitude less than that corresponding to sputtering. Therefore the contribution of the PSD mechanism to the moon's erosion can be considered negligible.

**The erosion rate due to micrometeoroid bombardment is estimated using the mass flow budgets constrained to the dust cloud measurements (Krueger et al., 2003). According to Table 5 in the paper of Krueger et al. (2003), the ratio of ejected to incident mass is $2x10^4$ and the escaping fraction of ejecta is equal to $2x10^{-4}$. Using the values contained in the work of Krueger et al. (2003), the escaping mass rate due to micrometeoroid bombardment is calculated to be of the order of 0.1 m/100Myr, which is less than that due to IS.**

**In the present study, effects like diversion of plasma flows by ionospheric electric fields (that could substantially lower surface fluxes) and acceleration of pickup ions (that could increase the total ion surface fluxes in some regions as suggested by Ip et al. (1998)) are not considered while estimating the erosion rates. Therefore, the above results are considered to be preliminary and a more detailed analysis of the erosion phenomenon, that will reflect possible differential Europa surface modifications, is intended in the future.**

In Europa, the vapour pressure produces the sublimation of a part of the released $H_2O$ molecules. Assuming a uniform temperature surface distribution, a sublimation rate of $10^{11}$ $H_2O$ $s^{-1}$ $m^{-2}$ is estimated (Shematovich et al., 2005). However, locally the sublimation rate can increase (e.g. at the dayside equatorial regions) up to $10^{15}$ $H_2O$ s$^{-1}$ m$^{-2}$ since the temperature at Europa's surface is not uniform (Spencer et al., 1999; Cassidy et al., 2007). However, we consider that the global average sublimating rate is more representative since, according to Shematovich et al. (2005), any localized region sublimating at this rate would be rapidly depleted of volatiles in the absence of continuously operating source.

On the basis of the above discussion, it is clearly understood that it is important to discriminate the active surface release process. As already mentioned in Section 2.2.1, the IS action can be distinguished from the PSD one since it produces also particles at higher energies and refractories. Hence, in order to discriminate the active mechanism it is desirable either to detect high energy particles (Orsini et al. 2009) by means of a neutral particle analyser or to detect the very weak component of the released refractories by means of a very sensitive and high-resolution mass spectrometer. The sputtered particles that have energies in the range between 10 eV and a few keVs are produced via the IS process, since in this energy window the fluxes released from all other processes can be considered negligible. In Fig. 8, the sputtered high energy (>10 eV) fluxes, produced by $S^+$ impinging ions impacting the Europa surface are presented. From comparison of Fig. 8 with Fig. 5 (left panel), it is clear that the IS action could be identified by the detection of energetic fraction of IS particles.

## 4. Conclusions

The goal of the present study is to investigate the effect of the various parameters on the surface ejection processes in Europa. We investigated the emission of $H_2O$ molecules released by the surface of the planet into its exosphere. Among all the processes considered, IS is the most effective process in ejecting molecules with sufficient energy to escape, as well as any refractory species that might be present (e.g., Si, Al, Mg). It is also found that contrary to IS and PSD, the IBS does not seem to contribute significantly to the total neutral particle release from Europa, in the energy range below 1 keV. Summarizing, our main results are the following:

1) The $H_2O$ density due to PSD ($6 \cdot 10^8$ $H_2O$ /m$^3$, on the surface of the illuminated side) is lower than that due to sputtering about 1.5 orders of magnitude. Furthermore, due to the variation of the Jupiter plasma precipitation, the IS can contribute locally, in different weights.

2) Considering that the particles that are emitted in the dayside and travel to the night side have to cross a longer path inside the atmosphere, the night-side density extracted from our model (neglecting the collisions) is probably overestimated. Depending on the mean free path assumed (ranging from 13 km to 78 km), one can estimate that only 0.3% to 3% of the particles are able to reach the night-side.

3) The most significant sputtered-particle flux and density come from the $S^+$ impinging ions and they are equal to 66% and 59% of total ones ($3.2 \cdot 10^{13}$ $H_2O$ /m$^2$ /s and $2.7 \cdot 10^{10}$ $H_2O$/m$^3$, respectively). These results are in general in good agreement with estimates obtained by other studies.

4) The neutral emission from the ESD process has a minor contribution to the total neutral density since the ESD-flux is 1.5 and 3 orders of magnitude less than that due to PSD and IS respectively. Moreover, the MIV process on Europa's surface, in general, can be considered of less significance since the energy input from meteoroids is of several orders of magnitude less than the particle energy fluxes (Cooper at el., 2001).

5) The total sputtering rate for Europa was calculated to be $2.0 \cdot 10^{27}$ $H_2O$ /s. The actual value of the total sputtering rate at some point on the planet's surface may exhibit variations, probably it is higher in the trailing face where the precipitation is modelled to be more intense (Paranicas et al., 2001).

6) **In the energy range below 1 keV,** the IBS process can be considered less important since it contributes much less than do the IS or the PSD**. In the higher energy range, above 1 keV, IBS dominates. Moreover, IBS is efficient for the light ions.**

7) The fraction of escaping particles via IS is 22% of all the sputtered ones, thus meaning a total rate of $4.4 \cdot 10^{26}$ $s^{-1}$, while the fraction of escaping particles via PSD is 30% of all the PSD-produced particles, thus meaning a total rate $3.3 \cdot 10^{24}$ $s^{-1}$. **This mean that the major agents for Europa's surface erosion are IS and the micrometeoroid bombardment on both the non-illuminated and illuminated side.**

A suggestion for defining the locally active release process is to discriminate the particle energy spectra and detect the high energy (>10 eV) particles of IS origin and /or to have a good mass spectrometer able to detect the low component of refractories (Milillo et al., 2005; Plainaki et al 2009). The flux of these high energy atoms **at Europa's** surface is calculated to be $5.2 \cdot 10^{12}$ /$m^2$/s.

At Europa, both the relevant laboratory yields and radiation fluxes have considerable uncertainties, and the source rate is highly variable due to temporal variability in the plasma conditions (Cassidy et al., 2007). Therefore, new in situ observations will be needed to separate the effect of surface reactions from the stimulating radiation flux. The results of this work could possibly be compared with some future observational data so that a better understanding of the active neutral particle release processes can be achieved.

In the future, a more precise neutral particle release model, taking into consideration, the following, is intended:
- existence of the tenuous atmosphere in collision regime
- effects on the sputtering, PSD and **sublimation** rate of a non uniform surface temperature
- the release rate due to minor species (e.g.: $H_2O_2$, $SO_2$, $CO_2$, etc...) on the moon surface
- **consideration of other processes that can result in producing also other molecules except $H_2O$ (e.g. $O_2$ via radiolysis)**

- a possible enhanced IS rate due to precursor radicals produced by UV irradiation
- **use of more accurate, possibly experimental, data for the modelling of neutral release processes (e.g. IBS yields)**

The present results can be considered as a first step to approach the other Jupiter's moon Ganymede for the similarity in the precipitating plasma composition and surface properties. This latter moon has a more complicated environment, due to the internal magnetic field that interacts with the Jupiter's magnetosphere and plasma. Finally, a comparison between the two moons' scenarios will give some light in the Jupiter's system investigation and evolution.

**Acknowledgments**

The authors thank both referees for their very useful comments and suggestions. The authors thank Giovanni Strazzulla from INAF – Osservatorio Astrofisico di Catania (Italy) for providing useful information, Chris Paranicas of the Applied Physics Laboratory of the John Hopkins University (USA) for useful comments on this manuscript and Robert. E. Johnson of the University of Virginia (USA) for important suggestions for improving the quality of this work. The authors also thank the Italian Space Agency for supporting their activities.

**Table-I**

Sputtering Input Parameters used in this analysis

| Parameter name | Symbol (unit) | Suggested Value | | | |
|---|---|---|---|---|---|
| Europa Radius | $R_E$ (km) | 1569 | | | |
| Europa Mass | $M_E$ (kg) | $4.8 \cdot 10^{22}$ | | | |
| Energy of the incident particle | $E_i$ (keV) | 100 | | | |
| Incident flux | $F_i$ (part/cm2s) | **$H^+$** | **$C^+$** | **$O^+$** | **$S^+$** |
| | | $1.5 \cdot 10^7$ Paranicas et al., 2002 | $1.8 \cdot 10^6$ Strazzulla et al., 2003 | $1.5 \cdot 10^6$ Paranicas et al., 2002 | $9 \cdot 10^6$ Paranicas et al., 2002 |
| Mass of the incident particle | $m_i$ (amu) | 1 | 12 | 16 | 32 |
| Sputtering Yield | Y (part/ion) | 6 Shi et al., 1995 | 10 Rocard, et al. (1986) | 50 Ip et al., 1997; 1998; Johnson, 1990; Shi et al., 1995 | 30 Johnson, 1990 |
| Mass of the ejected particle | $m_e$ (amu) | 18 | | | |
| Binding energy | $E_b$ (eV) | 0.05 (Boring et al., 1984; Haring et al., 1984) | | | |

# Figure Captions

*Fig.1 Energy distribution of the sputtered $H_2O$ particles emerging from the surface of Europa, in case of sulfur incident particles of 10 keV and 100keV, according to the formula given in Sigmund, 1969). Binding energy is assumed 0.054 eV. The vertical dashed line corresponds to the escape energy.*

*Fig. 2: Kinematic factor as a function of the scattering angle in case of O target atoms, for impinging $H^+$, $C^+$, $O^+$.*

*Fig. 3: Energy spectrum of the 10-keV protons back-scattered from an icy surface, as calculated with TRIM/SRIM software (Ziegler et al., 1988).*

*Fig.4 Probability distribution function for a PSD source ($T_e$=600K). The vertical dashed line corresponds to the escape energy*

*Fig.5: Upper panel: Sputtered-particles total $H_2O$ flux for different types of impinging ions of energy ~100 k eV ( the first panel corresponds to $H^+$, the second to $C^+$, the third to $O^+$, and the fourth one to $S^+$ ). The color-bar logarithmic scale is expressed in particles/s/$m^2$. Lower panel: Sputtered-particles total density for different types of impinging ions of energy ~100 k eV (the first panel corresponds to $H^+$, the second to $C^+$, the third to $O^+$ and the fourth one to $S^+$). The color-bar logarithmic scale is expressed in particles/$m^3$. Considered binding energy is 0.054 eV.*

*Fig.6: Flux (left panel) and density (right panel) of $H_2O$ particles emitted via PSD from the icy surface of Europa (logarithmic scale in particles/s/$m^2$ (left panel) and in particles/$m^3$ (right panel)).*

*Fig. 7: Intensity versus energy spectrum of the sputtered, back-scattered and PSD-ed neutrals.*

*Fig.8: Sputtered high energy fluxes particles/s/$m^2$ produced by $S^+$ impinging ions impacting Europa's surface, for effective binding energy equal to 0.05 eV.*

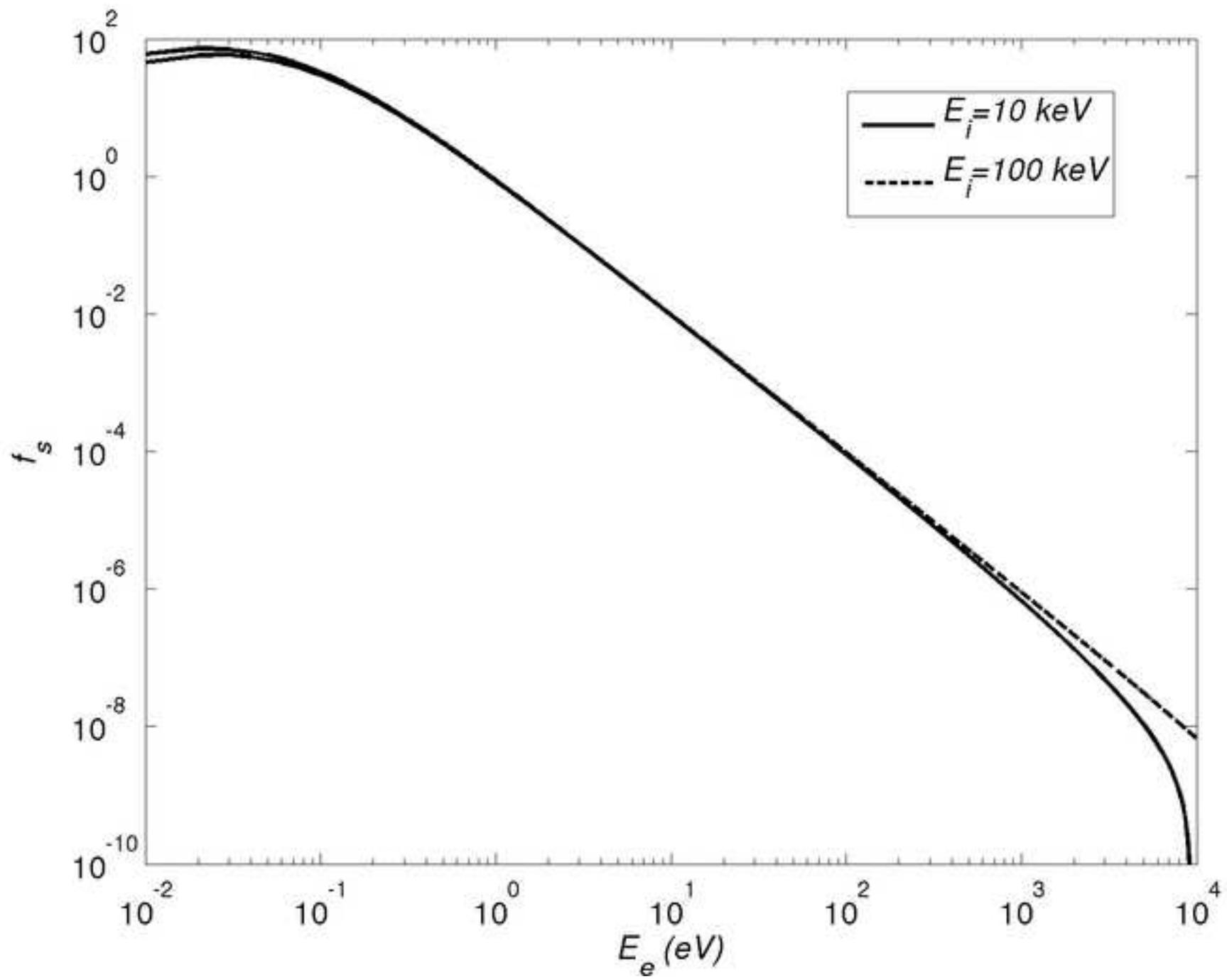

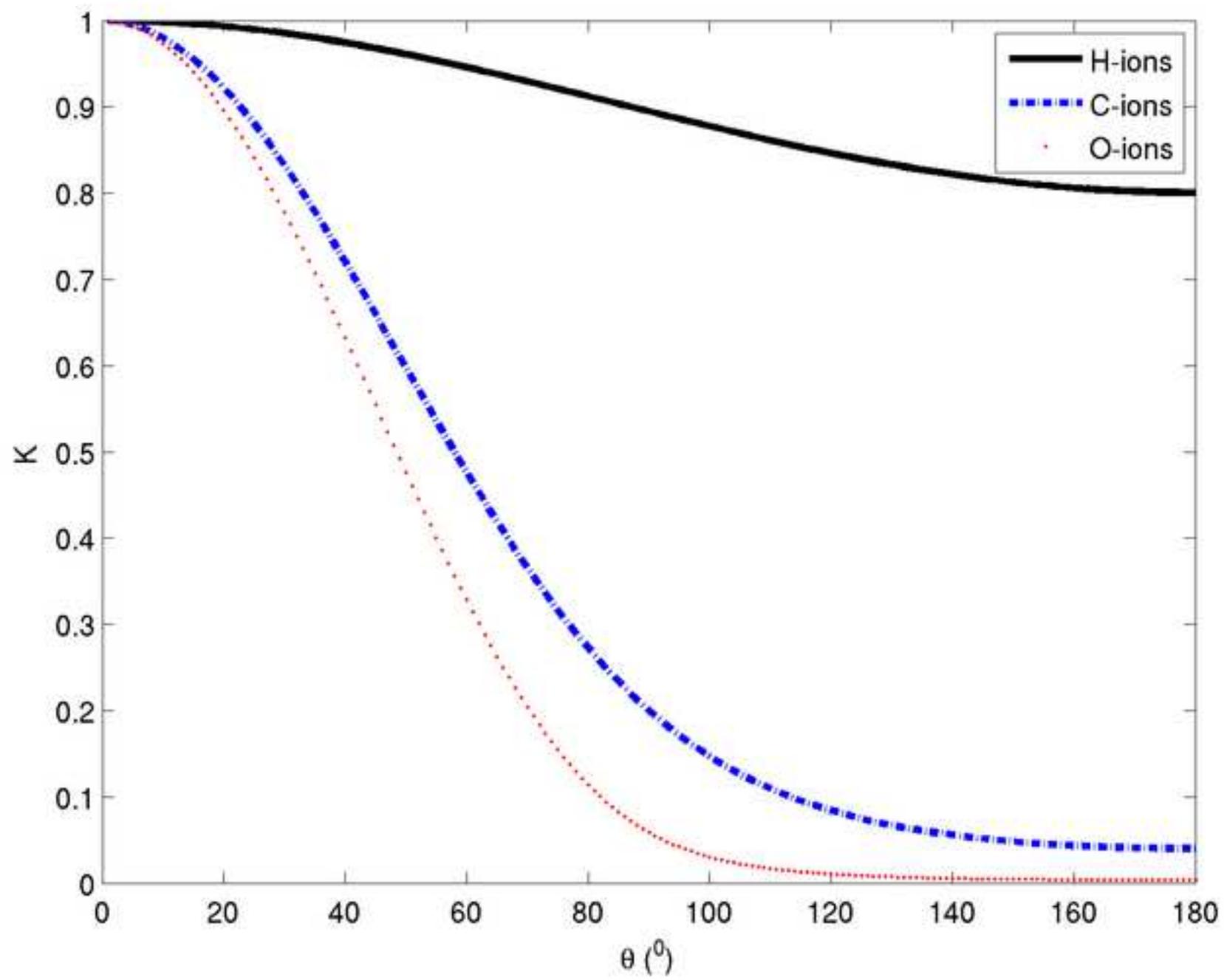

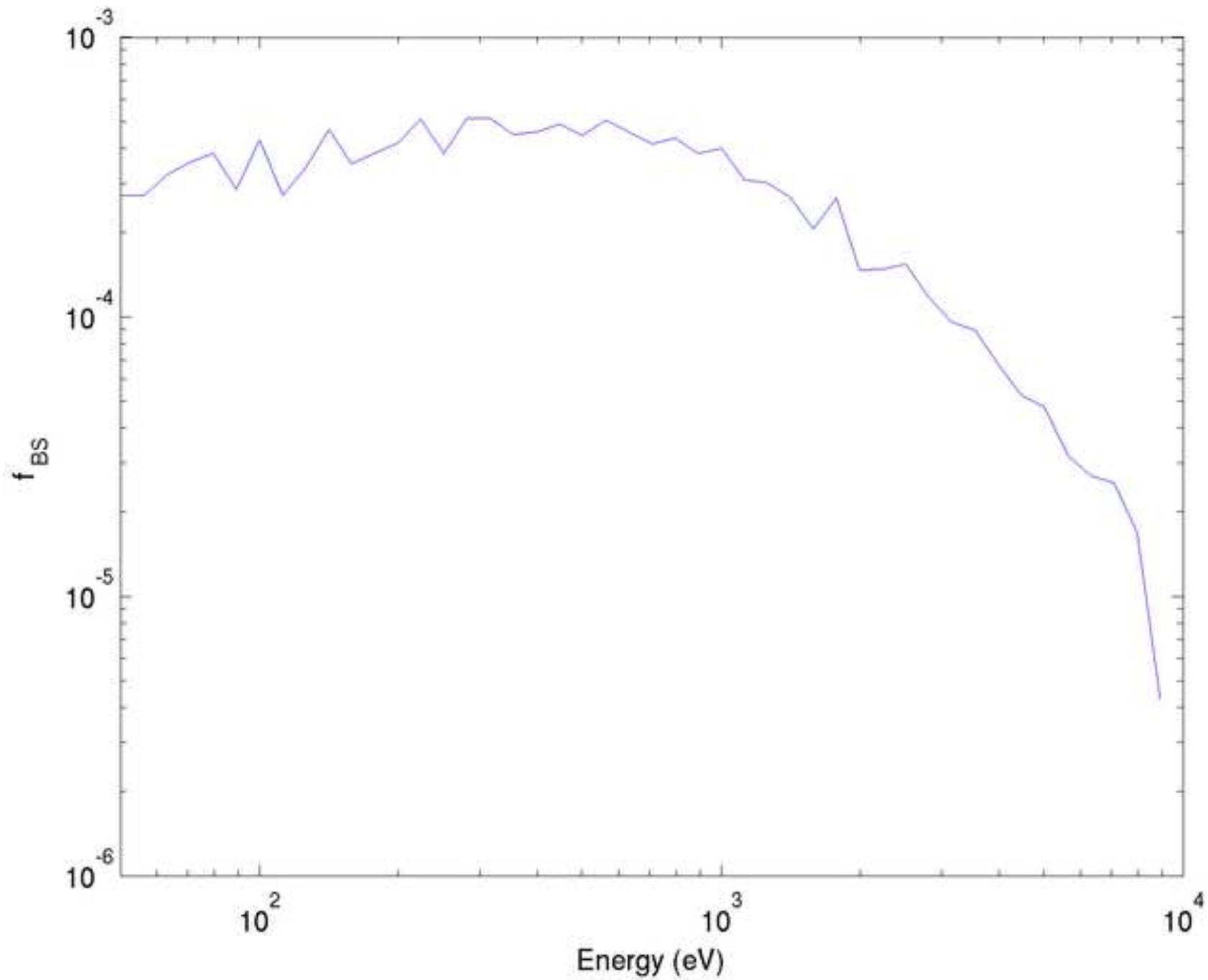

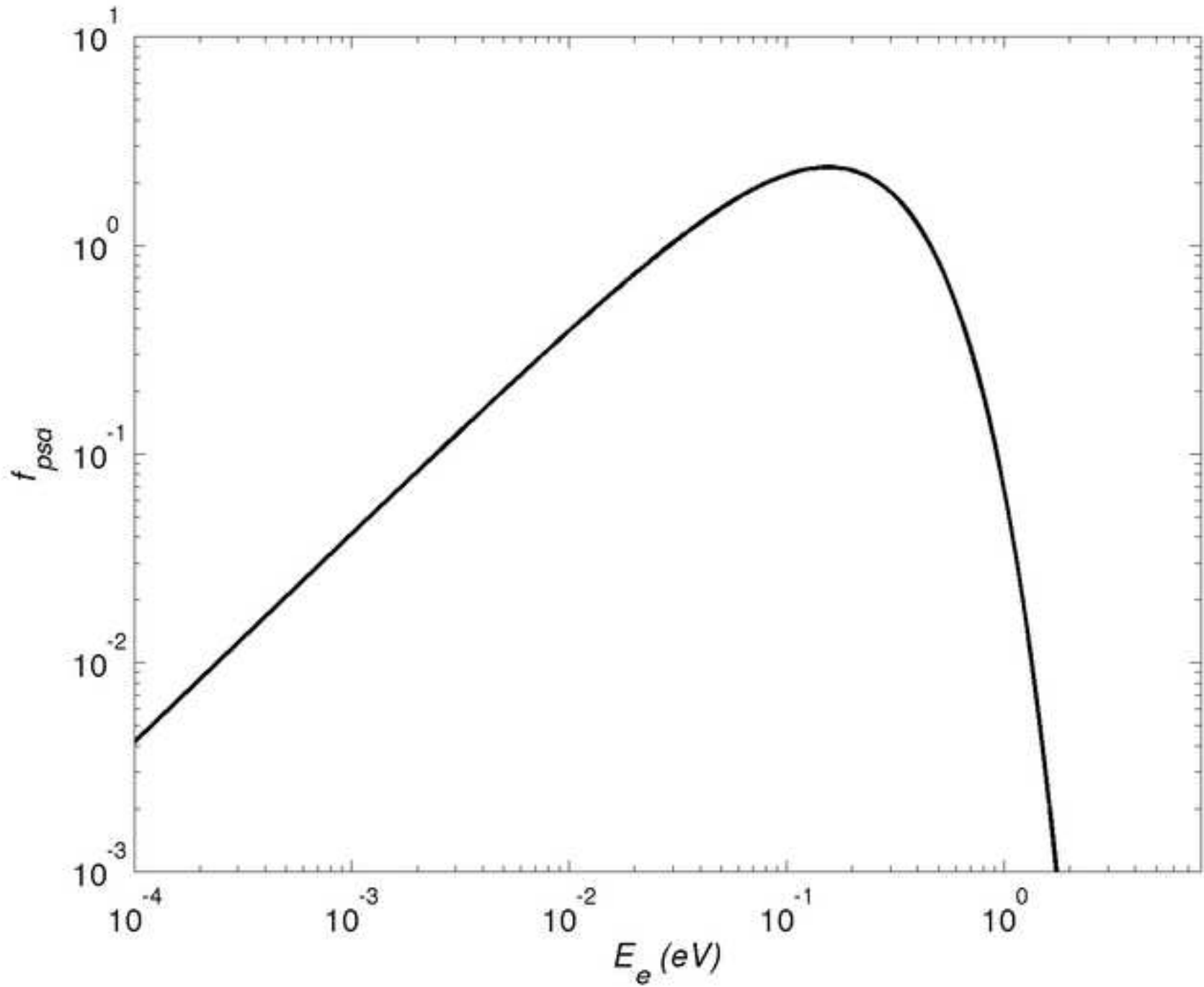

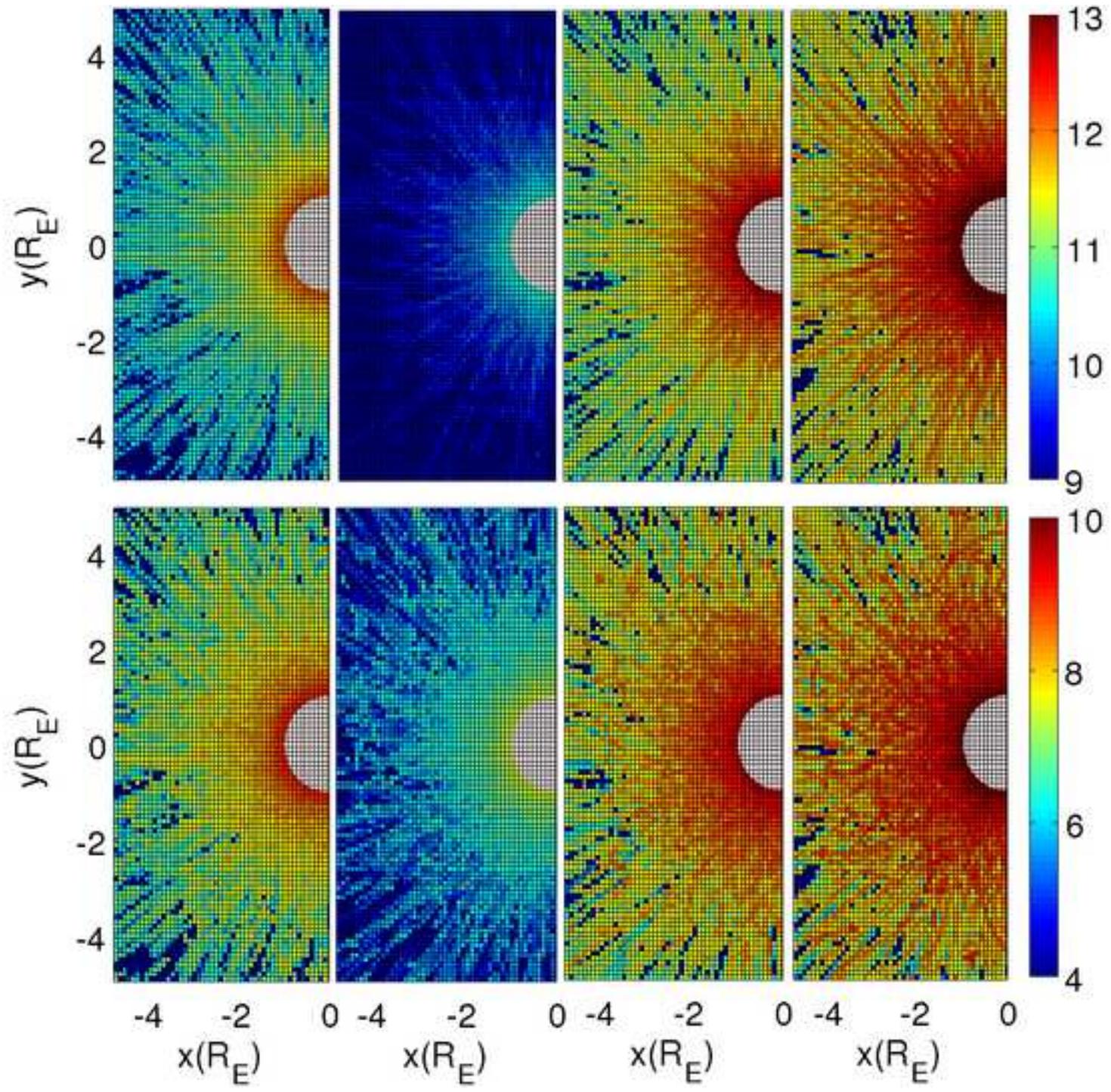

**2) Figure**
**Click here to download high resolution image**

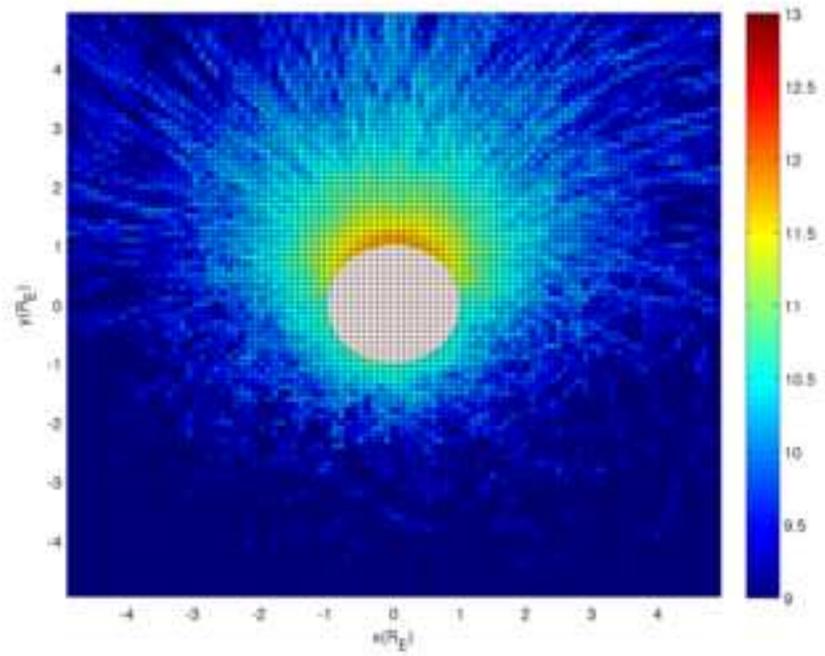 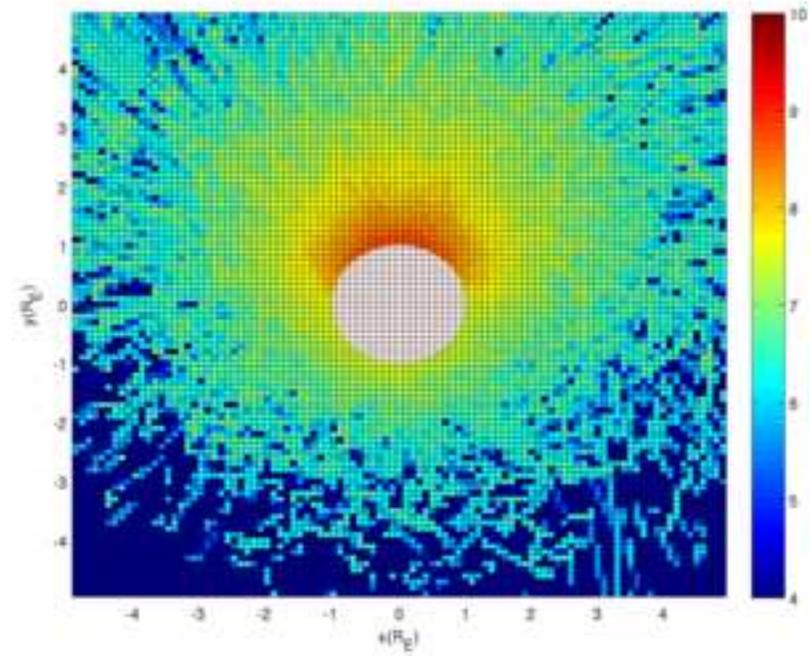

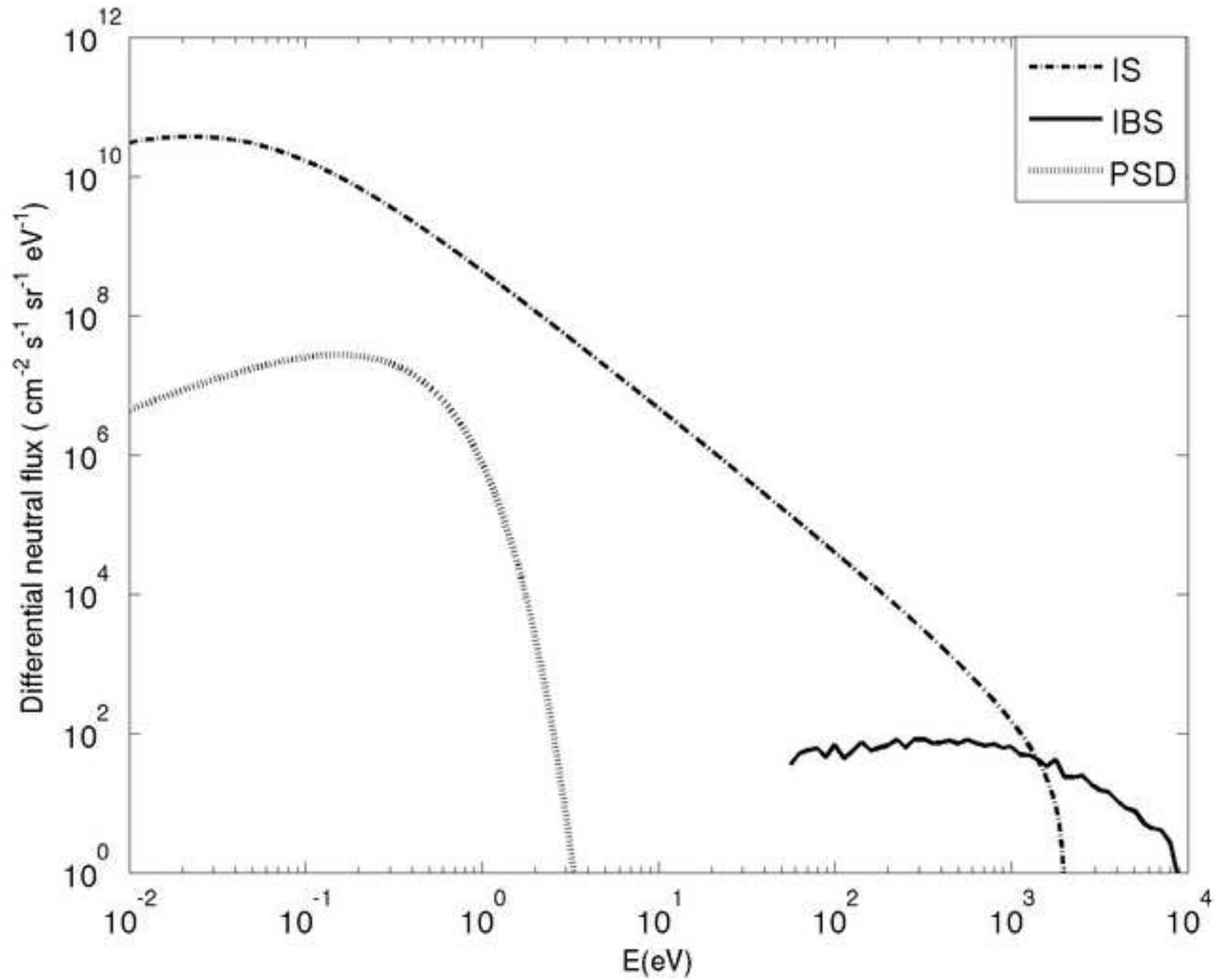

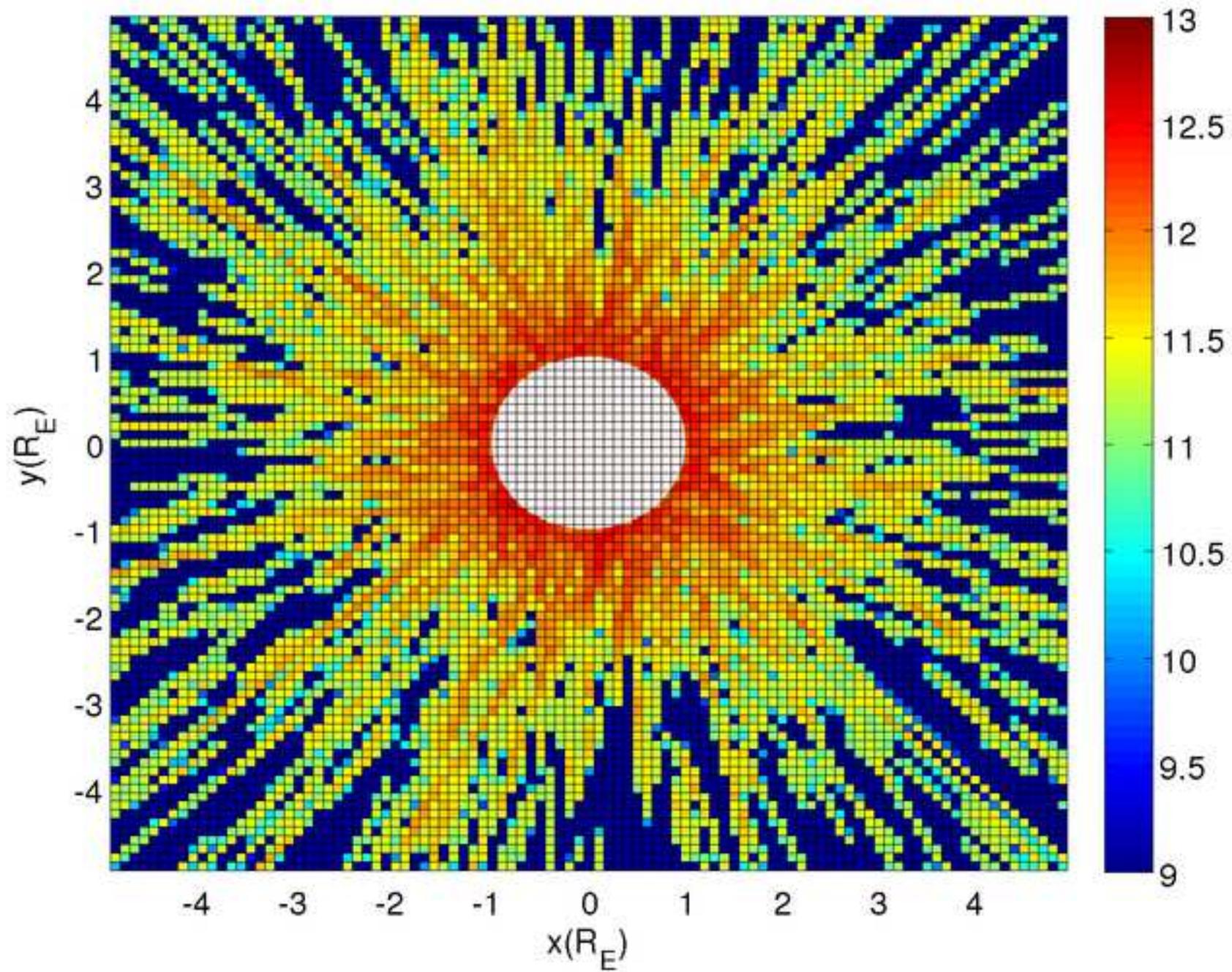